\documentstyle [preprint,aps,epsfig]{revtex}
\begin{document}
\tighten
\draft
\title{Superradiant Laser:\\
First-Order Phase Transition and Non-stationary Regime
}
\author{Christian Wiele\footnote{Electronic address:\
        christian.wiele@uni-essen.de} 
and Fritz Haake}
\address{Fachbereich Physik der Universit\"at-GH Essen, \\
  D-45117 Essen, Germany}
\author{Kazimierz Rz\c a\.zewski}
\address{
Centrum Fizyki Teoretycznej PAN and College of Science,\\
Al. Lotnik\'{o}w 32/46,
02-668 Warszawa, Poland
}
\maketitle
\begin{abstract}
We solve the superradiant laser model in two limiting cases. First the 
stationary low-pumping regime is considered where a first-order
phase transition in the semiclassical solution occurs. This discontinuity is 
smeared out in the quantum regime. Second, we solve the model in the 
non-stationary regime where we find a temporally periodic solution. For a
certain parameter range well separated pulses may occur. 
\end{abstract}

\pacs{PACS: 42.50.-p, 42.50.Fx, 42.50.Lc, 42.55.-f}

\section{introduction}
Superfluorescent pulses can be produced by $N$ collectively radiating 
identical atoms \cite{dicke54,bonifacio71,gross82} as these atoms decay 
from an initially excited state to the ground state. In contrast to such 
transient behavior would be the stationary output of the  
superradiant laser discussed in \cite{Haake93,Haake96,Haake96b}. 
Collective behavior would be manifested in the proportionality of the 
output intensity to $N^2$ and of the linewidth to $N^{-2}$. Moreover, as 
was shown in \cite{Haake93,Haake96} such a laser could display nearly 
perfect squeezing of the intensity fluctuations. 
The theory of a superradiant laser has up to now only been considered 
semiclassically in the stationary regime. As we shall show in the present
paper, the semiclassical solution needs some quantum mechanical corrections for a finite number of atoms and low pumping amplitudes. We also propose to extend
previous investigations towards a regime not allowing for a time independent stationary solution.

As in Ref. \cite{Haake93,Haake96} we 
consider the simplest model of a superradiant laser which accounts for $N$ 
three-level atoms placed inside a
resonator, Fig. \ref{FIG1}. We assume a resonant coherent two-photon pump process $0\to 2$. The two transitions
$2 \to 1$ and $1 \to 0$ are coupled to two resonant cavity modes $a,b$, which are
assumed to be so strongly damped that they are kept in adiabatic slavery by the atoms.
\\
The atoms are described by the collective population $(i=j)$ and polarisation
$(i\not= j)$ operators $S_{ij}=\sum_{\mu=1}^N
S_{ij}^\mu=\sum_{\mu}(|i\left>\right<j|)^\mu$.
We shall represent these atomic observables by  creation and annihilation operators
$z_i^\dagger,z_i$ with $S_{ij}=z_i^\dagger z_j$ and $[z_i,z_j^\dagger]=\delta_{ij}$; the two modes will be described by photon annihilation and creation operators $a, b, a^\dagger, b^\dagger$. The Hamiltonian $H_0$ for atoms and field modes reads
\begin{eqnarray}\label{hamiltonian}
H_0&=&-{\rm i}\hbar g_{12}(az_2^\dagger z_1-a^\dagger z_1^\dagger z_2) + \
{\rm i}\hbar g_{01}(bz_1^\dagger z_0 - b^\dagger z_0^\dagger z_1) \nonumber\\
&&{}+{\rm i}\hbar \Omega (z^\dagger_2z_0-z^\dagger_0z_2).
\end{eqnarray}
Here $\Omega$ is the amplitude of the external classical pump field, $g_{12}$ and
$g_{01}$ are the atom-field coupling constants for the transition 
$2\leftrightarrow 1$ and $1 \leftrightarrow 0$, respectively. 
As the modes are damped we have to add the two irreversible time rates of
change for the mode amplitudes $a,b$
\begin{equation}
\left(\frac{\partial a}{\partial t}\right)_{irr}=-\kappa_a a(t)+\sqrt{2\kappa_a}\eta_a(t),
\end{equation}
\begin{equation}
\left(\frac{\partial b}{\partial t}\right)_{irr}=-\kappa_b b(t)+\sqrt{2\kappa_b}\eta_b(t),
\end{equation}
where $\kappa_a, \kappa_b$ are the two damping constants and $\eta_a(t),
\eta_b(t)$ the corresponding quantum Langevin forces. The latter forces ensure
the preservation of the Bose commutators $[a(t),a^\dagger (t)]=[b(t),b^\dagger (t)]=1$. 
The Heisenberg-Langevin equations for the system with cavity damping are now
\begin{eqnarray}\label{heisenberg}
\dot{z_1^{\dagger}}&=&-g_{12} a z_2^{\dagger}+g_{01}b^{\dagger}z_0^{\dagger},\\
\dot{z_0^{\dagger}}&=&-\Omega z_2^{\dagger}-g_{01}b z_1^{\dagger},\\
\dot{z_2^{\dagger}}&=&\Omega z_0^{\dagger}+g_{12} a^{\dagger} z_1^{\dagger},\\
\dot{a}&=&-g_{12}z_1^{\dagger}z_2 - \kappa_a a +\sqrt{2\kappa_a}\eta_a,\\
\label{heis2}
\dot{b}&=&-g_{01}z_0^{\dagger}z_1 - \kappa_b b +\sqrt{2\kappa_b}\eta_b.
\end{eqnarray}
These equations have been solved before in \cite{Haake96} under some
assumptions that we shortly recapitulate here. First the $b$-mode is
taken to be strongly damped so it can be adiabatically eliminated. For later use
we define the two damping constants $\gamma_a,\gamma_b$:
\begin{equation}
\gamma_a=\frac{g_{12}^2}{\kappa_a}, \quad \gamma_b=\frac{g_{01}^2}{\kappa_b}.
\end{equation}
Then an analytic solution can be given in the semiclassical limit $N\gg1$ in
which the observables are represented as a sum $X=\bar{X}+\delta X$ of a dominant
classical term $\bar{X}\sim N$ and a "small" operator valued fluctuation
$\delta X$.
The classical term is evaluated in the stationary regime by dropping
$\dot{X}$ and the Langevin forces. The resulting solutions can be expressed in
terms of a dimensionless coupling strength $c$ and an effective pump strength $p$: 
\begin{equation}
  \label{eq:pc}
   c=\frac{\gamma_a}{\gamma_b}, \quad p=\frac{\Omega}{N\sqrt{c}\gamma_b}.
\end{equation}
The solution for the level populations and the field amplitude is    
\begin{eqnarray}
  \label{eq:semi}
  \bar{S}_{00}&=&\frac{Nc(1-p)}{1+c}, \quad \bar{S}_{11}=Np,\quad\nonumber\\
  \bar{S}_{22}&=&\frac{N(1-p)}{1+c}, \quad \
\bar{a}=\frac{N\gamma_b}{g_{12}}\frac{c\sqrt{p(1-p)}}{\sqrt{1+c}},
\end{eqnarray}
which is stable under the constraints $0\leq p \leq 1$ and $c>1$ in the case
we are interested in (additional adiabatic elimination of the $a$-mode).  
Notice that for $p\to 0$ and $p\to1$ some of the observables vanish. This is
in conflict to the assumption of a large mean value and small fluctuations, so
the semiclassical approximation will break down at these points. Moreover,
this solution cannot be correct in the limit $p\to 0$ since it does not correspond 
to the ground state $\left<S_{00}\right>=N$. It is physically clear, however, 
that without pumping all atoms will eventually settle 
in the lowest state, due to damping of the levels $2$ and $1$. 
To improve on the semiclassical prediction of a discontinuity of $\langle S_{22}\rangle$ 
for $p=0$ we have to consider the full quantum mechanical 
solution. This will 
be done in two ways. First, we develop a perturbative small-$p$
expansion. Second, we give a recurrence relation for the whole stationary
density operator $\rho$ which can be evaluated numerically. \\
In the final section of this paper we shall take a look at the regime where
(\ref{eq:semi}) is not a stable solution anymore. 

\section{First-order Phase Transition in the stationary regime}
For the discussion of the quantum mechanical solution we will switch to the
master equation of the superradiant laser. Additionally we
assume the $a$-mode also to be so strongly damped that we can eliminate it adiabatically.
The master equation then reads
\begin{equation}
  \label{eq:master1}
  \dot{\rho}=(L_{02} + L_{21} + L_{10})\rho
\end{equation}
with
\begin{eqnarray}
  \label{eq:master3}
  L_{02}\rho&=&\Omega[S_{02}-S_{20},\rho],\\
  L_{21}\rho&=&\gamma_a\{[S_{12},\rho S_{21}]+[S_{12}\rho,S_{21}]\},\\
  L_{10}\rho&=&\gamma_b\{[S_{01},\rho S_{10}]+[S_{01}\rho,S_{10}]\}.
\end{eqnarray}
For the discussion we expand $\rho$ in the fully symmetric and normalized
states $|1^m;2^l\rangle$ with $l$ atoms in level 2, $m$
atoms in level 1, and $N-m-l$ atoms in level 0. The short-hand notation
$|0\rangle\equiv|1^0;2^0\rangle$ is used for the ground state.
\subsection{Second-Order Perturbation Expansion}
In this subsection we take a closer look at the neighborhood of the point
$p=0$. For zero pumping the atoms will eventually all settle in the ground
state, so the stationary density operator is 
$\bar{\rho}(p=0)=|0\rangle\langle0|$. For small non-zero pumping $\bar{\rho}$ may be
expanded in a series
\begin{equation}
  \label{eq:rhoseries}
  \bar{\rho} = \rho^{(0)} + \rho^{(1)} + \rho^{(2)} + ...  
\end{equation}
where $\rho^{(n)}\propto p^n$. With $\lambda\equiv L_{21}+L_{10}$ the terms up to 
second order are
\begin{equation}
  \label{eq:0th}
  \rho^{(0)}= |0\left>\right<0|,
\end{equation}
\begin{eqnarray}
  \label{eq:1st}
  \rho^{(1)}&=&\lim_{t \to \infty}\int\limits_0^t dt_1\, {\rm e}^{\lambda (t-t_1)} \
  L_{02}{\rm e}^{\lambda t_1} |0\left>\right<0|\nonumber\\
  &=&-\frac{pN^{3/2}}{\sqrt{c}} (|2^1\left>\right<0|+|0\left>\right<2^1|),
\end{eqnarray}
\begin{eqnarray}
  \label{eq:2nd}
  \rho^{(2)}&=&\lim_{t \to \infty}\int\limits_0^t dt_1\, {\rm e}^{\lambda (t-t_1)} \
  \int\limits_0^{t_1} dt_2 \nonumber\\
&&\qquad\times L_{02}{\rm e}^{\lambda (t_1-t_2)}L_{02}{\rm e}^{\lambda t_2}|0\left>\right<0|\nonumber\\
  &=&\left(1-p^2N^2\frac{N+c}{c}\right)|0\left>\right<0|\nonumber\\
  &&{}+\frac{p^2N^3}{c}|2^1\left>\right<2^1|\
  +p^2N^2|1^1\left>\right<1^1|\nonumber\\
  &&{}+\frac{p^2N^3}{c}\sqrt{\frac{N-1}{2N}}\left(|2^2\left>\right<0|+|0\left>\right<2^2|\right).
\end{eqnarray}
These entail the mean occupation numbers  
\begin{eqnarray}
  \label{eq:perturbation}
  \left<S_{00}\right>&=&N-p^2N^2\frac{N+c}{c}+{\cal O}(p^3),\\ 
  \left<S_{11}\right>&=&p^2N^2+{\cal O}(p^3),\\
  \left<S_{22}\right>&=&\frac{p^2N^3}{c}+{\cal O}(p^3).
\end{eqnarray}
Starting from the ground state the corrections are proportional to $p^2$
and $\langle S_{22} \rangle$ is of order $N$ (as the semiclassical solution) 
for $p=1/N$. Thus the discontinuity of the semiclassical solution is smeared
out over a range $0 \leq p \leq 1/N$. Further calculations show that the
fluctuations of the observables vanish for $p\to0$. All of this is 
reminiscent of a first-order phase transition.
\subsection{Recurrence Relation}
Having studied the perturbation expansion we now prepare to look for a
complete quantum mechanical description. To that end we again expand the 
density operator $\rho$ in the states $|1^m;2^l\rangle$ introduced above. 
One may check that coherences with respect to different numbers of atoms in 
level 1 decay to zero. We take this into account by the simple ansatz 
\begin{equation}
  \label{eq:rho}
  \bar{\rho}=\sum_{0\leq m+r\leq N \atop 0 \leq m+l \leq N}
  \bar{\rho}_{l,m,r}|1^m;2^l\left. \right> \left< \right. 1^m;2^r |.
\end{equation}
This expansion is inserted into (\ref{eq:master1}) and we find a recurrence relation
for the stationary solution $\bar{\rho}_{l,m,r}$
\begin{eqnarray}
  \label{eq:recurrence} 
    0&=&pN\sqrt{c}\left\{ \sqrt{(N-m-l+1)l}\bar{\rho}_{l-1,m,r}\right. \nonumber\\
      &&{}-\sqrt{(N-m-l)(l+1)}\bar{\rho}_{l+1,m,r} \nonumber\\
      &&{} +\sqrt{(N-m-r+1)r}\bar{\rho}_{l,m,r-1}\nonumber \\
      &&\left. {}-\sqrt{(N-m-r)(r+1)}\bar{\rho}_{l,m,r+1}\right\} \nonumber\\
    &&{}+c\left\{2m\sqrt{(r+1)(l+1)}\bar{\rho}_{l+1,m-1,r+1}\right. \nonumber \\
    &&{}\left. -(m+1)(l+1)\bar{\rho}_{l,m,r}\right\} \nonumber\\
    &&{}+2(m+1)\sqrt{(N-m-l)(N-m-r)}\bar{\rho}_{l,m+1,r}\nonumber\\
    &&{}-m(2N-2m+2-l-r)\bar{\rho}_{l,m,r}.
\end{eqnarray}
We have not managed to find an analytical solution to this equation, but have 
solved it numerically. Since the number of involved variables is roughly proportional 
to $N^3$, only up to 30 atoms have been considered. The results are shown in 
Fig. \ref{FIG2}. Beside the corrections at $p=0$ one can make out that the
semiclassical solution has to be corrected for $p\to 1$ as well, as explained
above. In the intermediate $p$ range the semiclassical result is quite good,
even for the moderate values of $N$ studied.
Since the transition from quantum to semiclassical behavior can already be
seen in our calculations, it seems not necessary to go to higher $N$.

\section{Non-stationary regime}
In this section we consider the non-stationary solution of our system in
the bad-cavity limit. As indicated in \cite{Haake96b} we expect to find a
temporally periodic behavior of the laser field. So we set $c<1$, where the 
semiclassical stationary solution is not stable. Starting from the 
Heisenberg-Langevin equations (\ref{heisenberg})-(\ref{heis2}), we assume 
strong, saturated driving between levels 0 and 2,
\begin{eqnarray}
z_2&=&\sqrt{N}\sin (\Omega t),\\
z_0&=&\sqrt{N}\cos (\Omega t).
\end{eqnarray}
With the assumption of very fast relaxing cavity modes we can once more eliminate
$a$ and $b$ adiabatically,
\begin{equation}
a(t)=-\frac{g_{12}\sqrt{N}}{\kappa_a}\sin (\Omega t) z_1^{\dagger} + \
 \sqrt{\frac{2}{\kappa_a}}\eta_a(t),
\end{equation}
\begin{equation}
b^{\dagger}(t)=-\frac{g_{01}\sqrt{N}}{\kappa_b}\sin (\Omega t) z_1^{\dagger} + \
 \sqrt{\frac{2}{\kappa_b}}\eta^{\dagger}_b(t),
\end{equation}
and find a linear equation for $z_1^\dagger$
\begin{equation}\label{z1dot}
\dot{z_1^{\dagger}}=\left[\frac{\Gamma_a-\Gamma_b}{2}-\
\frac{\Gamma_a+\Gamma_b}{2}\cos (2\Omega t)\right] z_1^{\dagger} + A(t),
\end{equation}
\begin{equation}
\Gamma_a=\frac{g_{12}^2 N}{\kappa_a}, \quad \Gamma_b=\frac{g_{01}^2 N}{\kappa_b},
\end{equation}
\begin{equation}
A(t)=-\sqrt{2\Gamma_a}\sin (\Omega t) \eta_a(t) + \
\sqrt{2\Gamma_b} \cos(\Omega t)\eta_b^{\dagger}(t).
\end{equation}
Now equation (\ref{z1dot}) is easily solved
\begin{equation}
z_1^{\dagger}(t)={\rm e}^{u(t)}z_1^{\dagger}(0)+\
{\rm e}^{u(t)}\int\limits_0^t{\rm e}^{-u(t')}A(t')dt'
\end{equation}
with
\begin{equation}
u(t)=\frac{\Gamma_a-\Gamma_b}{2}t-\frac{\Gamma_a+\Gamma_b}{4\Omega}\sin(2\Omega t).
\end{equation}
Clearly we encounter a periodic behavior with period $1/2\Omega$.
We use this as time scale and define the dimensionless parameters 
$s,d,$ and $\tau$ 
\begin{eqnarray}
s&=&\frac{\Gamma_b+\Gamma_a}{2\Omega}=\frac{1+c}{2p\sqrt{c}},\\
d&=&\frac{\Gamma_b-\Gamma_a}{2\Omega}=\frac{1-c}{2p\sqrt{c}},\\
\tau&=&2\Omega t.
\end{eqnarray}
From this the mean number of photons in the $a$-mode is 
\begin{eqnarray}
\left<a^{\dagger}(t)a(t)\right>&=&\
\frac{\Gamma_a\Gamma_b}{\kappa_a\Omega}\sin^2(\tau/2)\
{\rm e}^{-d\tau-s\sin(\tau)}\nonumber\\
&&\qquad\times\int\limits_0^\tau \
\cos^2(\tau'/2){\rm e}^{d\tau'+s \sin(\tau')}d\tau'\nonumber\\
&\equiv&\frac{\Gamma_a\Gamma_b}{\kappa_a\Omega}{\rm Int}(\tau).
\end{eqnarray}
The time integration may be performed after expansion in terms of the
Bessel functions $I_n(s)$,
\begin{equation}
{\rm e}^{s\sin(\tau)}=\sum\limits_{n=-\infty}^{+\infty}\
(-{\rm i})^nI_n(s){\rm e}^{{\rm i}n\tau},
\end{equation}
and after the death of initial transients ${\rm e}^{-d\tau}\to 0$ we get
\begin{eqnarray}
{\rm Int}(\tau)&\to&\frac{1}{2}\sin^2(\tau/2){\rm e}^{-s\sin(\tau)}\nonumber\\
&&\qquad\times\sum\limits_{n=-\infty}^{+\infty}\
\frac{{\rm e}^{{\rm i}n\tau}}{d+{\rm i}n}(-{\rm i})^nI_n(s)\left(1+{\rm i}\frac{n}{s}\right).
\end{eqnarray}
To illustrate the temporal periodicity of this solution we give some examples  for various parameters $p,c$ in Fig. \ref{FIG3}.  
For the limiting case $s\gg1$, which implies small $p$, we find well separated
pulses the width and height of which may be estimated in a Gaussian 
approximation. With the parameter $\epsilon=d/s$ we find
\begin{eqnarray}
{\rm Int}(\tau)&=&\frac{1}{4}\frac{1}{s^2p^2}\sqrt{2\pi p}\
{\rm e}^{[2/p-d(\tau_{max}-\tau_{min})]}\cdot\nonumber\\
&&\cdot {\rm e}^{-(\tau-\tau_{max})^2/2p}
\end{eqnarray}
with 
\begin{eqnarray}
\cos(\tau_{max})&=&-\epsilon, \qquad \sin(\tau_{max})=-\sqrt{1-\epsilon^2},\\
\cos(\tau_{min})&=&-\epsilon, \qquad \sin(\tau_{min})=+\sqrt{1-\epsilon^2}.
\end{eqnarray}
So the pulse width is 
\begin{equation}
\sigma=\sqrt{p}=\sqrt{\frac{\Omega\sqrt{\kappa_a\kappa_b}}{Ng_{12}g_{01}}}.
\end{equation}
Due to the assumptions of saturated driving and strongly damped cavity
modes we have $\Omega\gg g_{12},g_{01}$, $\kappa_a\gg g_{12}$, and  
$\kappa_b\gg g_{01}$. This entails the need for a large number of atoms to 
make $\sigma$ small, i.e. to get narrow and well separated pulses. 

The time-dependent spectrum of a light signal which accounts for the width of
the filter $\Gamma$ is given as \cite{Eberly76}:
\begin{eqnarray}
S_\Gamma(\omega,t)&=&\int\limits_0^tdt_1\int\limits_0^tdt_2
{\rm e}^{-\Gamma(2t-t_1-t_2)}\
{\rm e}^{{\rm i}\omega(t_1-t_2)}\nonumber\\
&&\qquad\times\langle a^\dagger(t_1) a(t_2)\rangle
\end{eqnarray}
Since we have explicit expressions for the $a,a^\dagger$ operators, we plug
them in and after some integrations sending the width of the filter 
$\Gamma\to 0$ and the observation time $t \to \infty$ we get the following
time-averaged expression
\begin{eqnarray}
S(\omega)&\to&\frac{\Gamma_a\Gamma_b}{16\Omega^3}\sum\limits_n \sum\limits_l \
\frac{(-1)^n I_n(s)\left(1+\frac{{\rm i}n}{s}\right)}{d+{\rm i}n}\nonumber\\
&&\qquad \times \left[ I_{n+l+1}\left(\frac{s}{2}\right) - \
  {\rm i} I_{n+l}\left(\frac{s}{2}\right)\right]  \nonumber \\
&&\qquad \times \left[ I_{l+1}\left(\frac{s}{2}\right) + \
  {\rm i} I_{l}\left(\frac{s}{2}\right)\right]  \nonumber \\
&&\times\left[\frac{1}{d+{\rm i}(2n+2l+1+\omega/\Omega)}\right. \nonumber\\
&&{}+\left. \frac{1}{d+{\rm i}(2n+2l+1-\omega/\Omega)}\right].
\end{eqnarray}
We have used the same parameters $p,c$ as in Fig. \ref{FIG3} to illustrate the
spectra of some of these pulses in Fig. \ref{FIG4}. Of course high harmonics of the
Rabi frequency $\Omega$ are enhanced for $\sigma\to0$.

\section{Acknowledgements}
One of us (K.R.) would like to thank the colleagues at the {\it Fachbereich Physik}
in Essen for their hospitality and the {\it Humboldt Foundation} for
financial support. Support of the Sonderforschungsbereich {\it "Unordnung und 
gro{\ss}e Fluktuationen"} is also gratefully acknowledged. We are deeply indebted 
to Mikhail I. Kolobov, Elisabeth Giacobino, Claude Fabre, and Serge Reynaud 
for their invaluable contributions to this project.


\begin{figure}
\caption[FIG1]{Scheme of the 3-level superradiant laser.}
\label{FIG1}
\end{figure}
\begin{figure}
\caption[FIG2]{Occupation numbers of the three levels derived through
the recurrence relation. Even for our moderate number of atoms one can clearly 
see the transition from the quantum to the semiclassical regime. The 
discontinuity of $\langle S_{22}\rangle$ for $p\to0$ in the semiclassical 
solution is smeared out over a range $1/N$.}
\label{FIG2}
\end{figure}
\begin{figure}
\caption[FIG3]{The non-stationary pulsed solution for different parameters. 
As indicated by the calculations the pulse width is given by 
$\sqrt{p}$ and is independent of $c$.}
\label{FIG3}
\end{figure}
\begin{figure}
\caption[FIG4]{Spectra of the pulsed solution for different parameters. As one
expects the narrow pulse ($p=0.1$) has higher harmonics than the broad pulse 
($p=2$).}
\label{FIG4}
\end{figure}

\clearpage
\onecolumn
\begin{figure}
\begin{center}
\epsfig{file=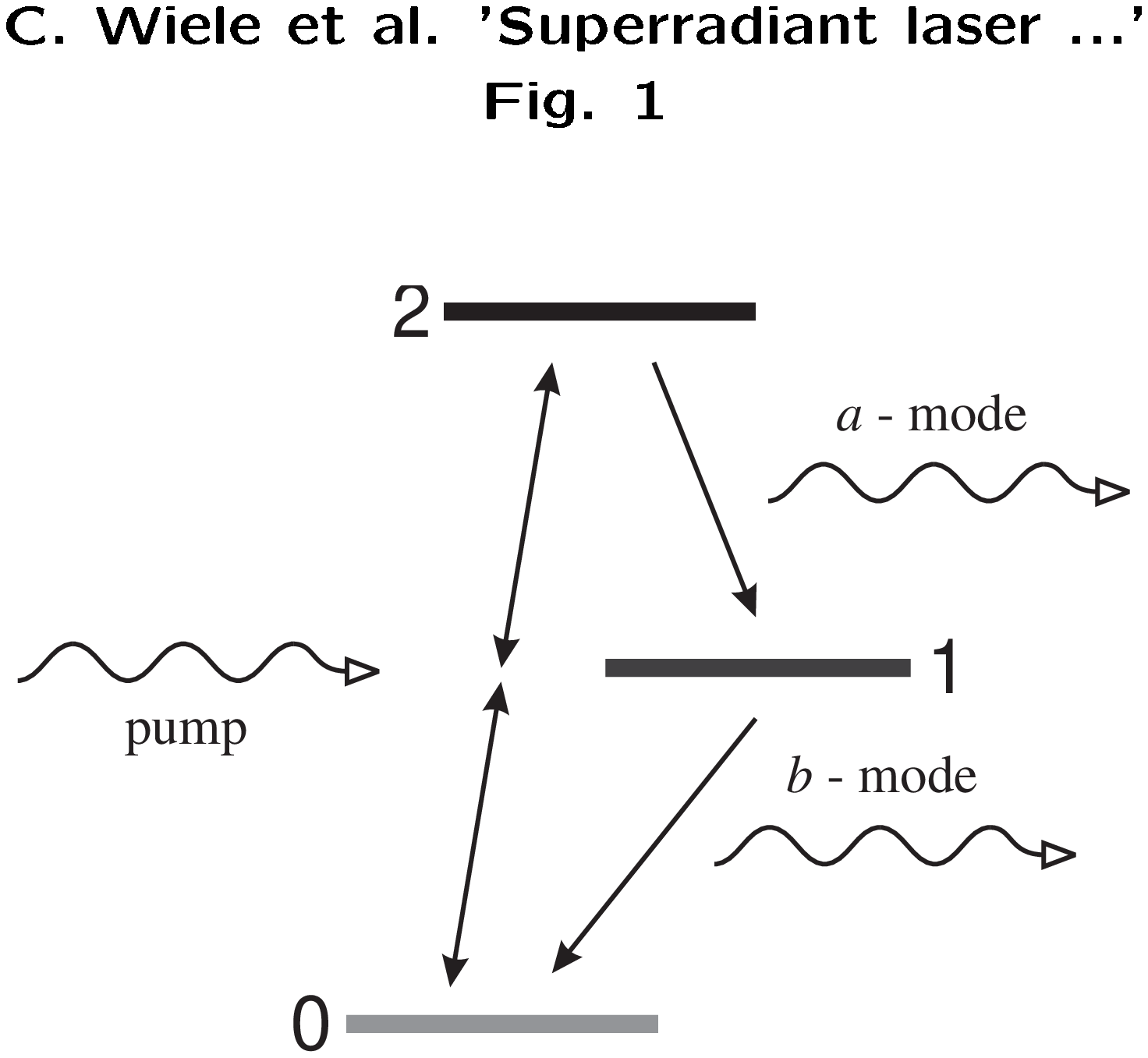,width=16truecm}
\end{center}
\end{figure}

\begin{figure}
\begin{center}
\epsfig{file=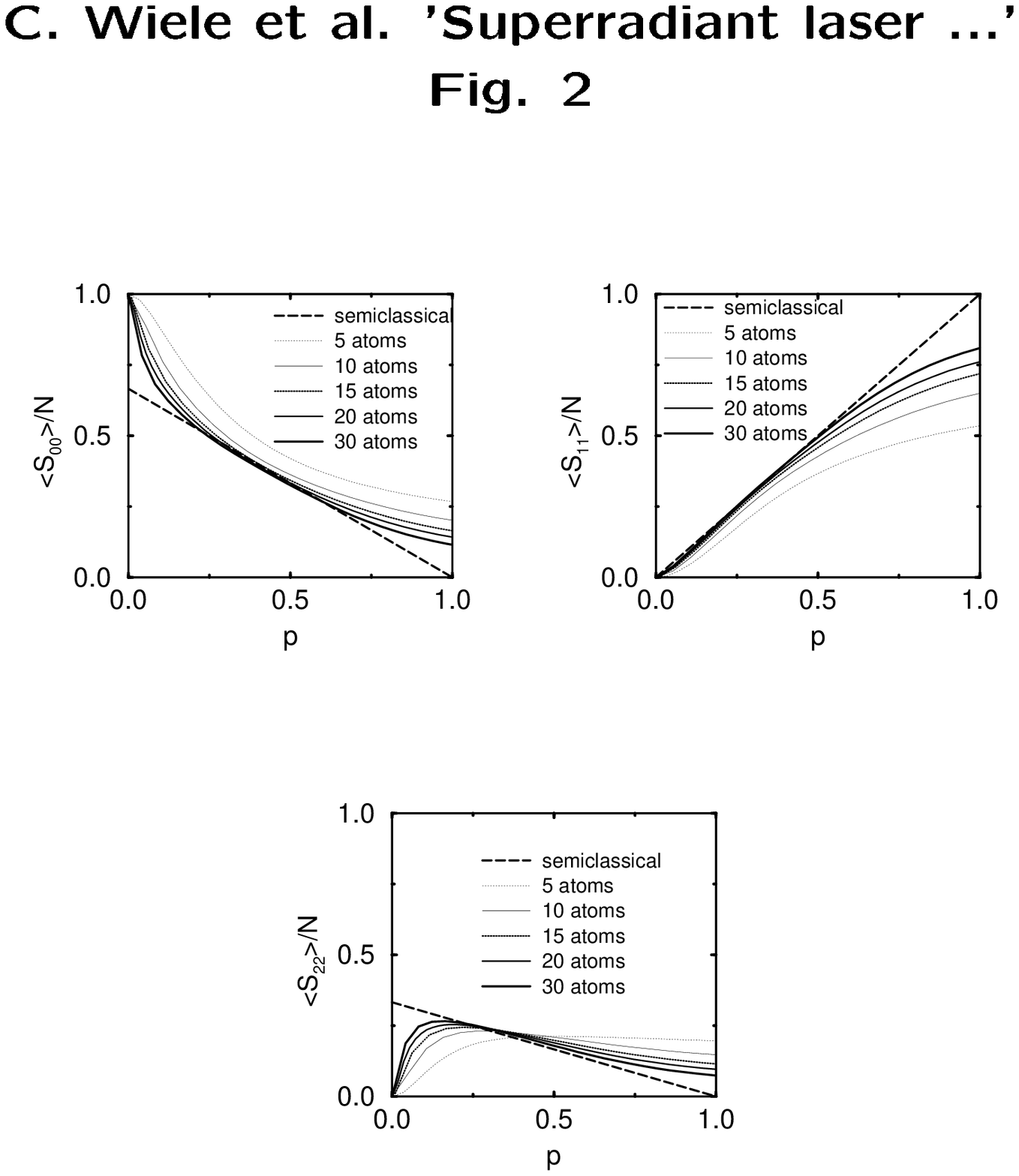,width=16truecm}
\end{center}
\end{figure}

\begin{figure}
\begin{center}
\epsfig{file=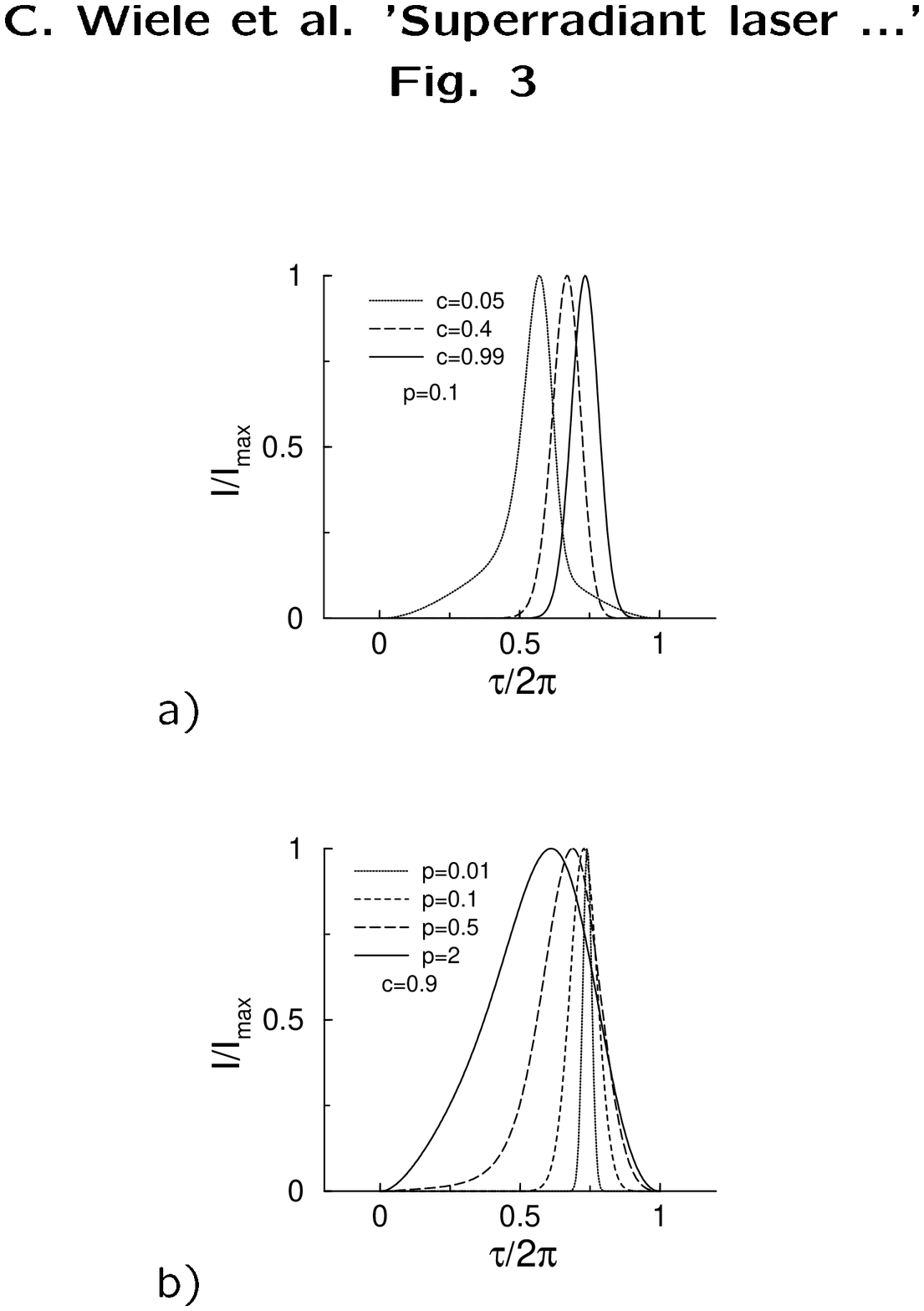,width=16truecm}
\end{center}
\end{figure}

\begin{figure}
\begin{center}
\epsfig{file=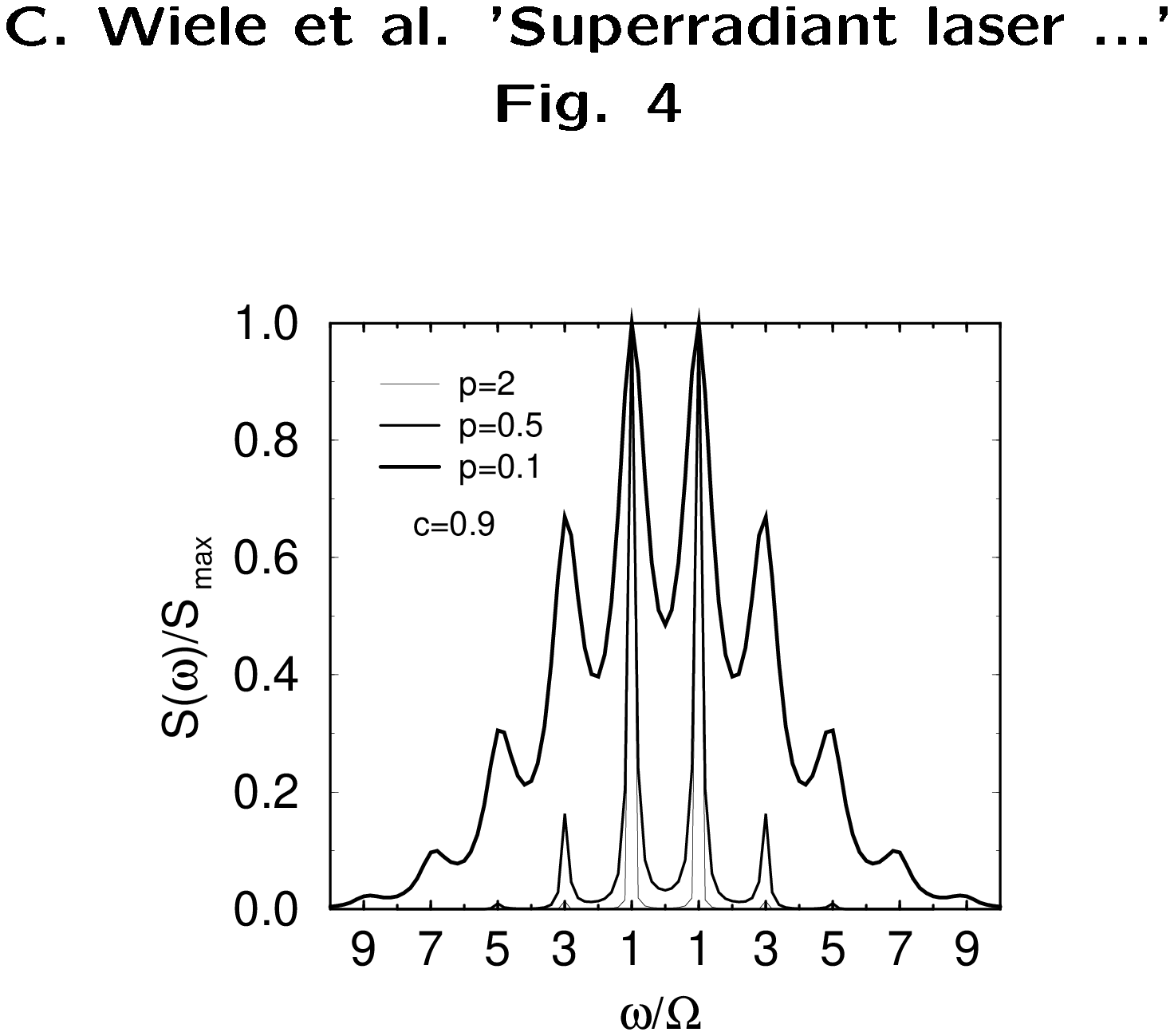,width=16truecm}
\end{center}
\end{figure}
\end{document}